\documentclass[cameraready]{Interspeech}

\usepackage{colortbl,xcolor}
\usepackage[dvipsnames]{xcolor} 
\usepackage{multirow} 
\usepackage{cite}
\usepackage{url}

\title{ERM-MinMaxGAP: Benchmarking and Mitigating Gender Bias in Multilingual Multimodal Speech-LLM Emotion Recognition}

\author[affiliation={1}, orcid=0009-0003-6650-8982]{Zi Haur}{Pang}
\author[affiliation={2}, orcid=0000-0003-1920-5228, correspondingauthor]{Xiaoxue}{Gao}
\author[affiliation={1}, orcid=0000-0002-2686-2296]{Tatsuya}{Kawahara}
\author[affiliation={2}, orcid=0000-0003-0872-5877]{Nancy F.}{Chen}

\address{
    $^1$ Kyoto University, Japan \\
    $^2$ Agency for Science, Technology, and Research (A*STAR), Singapore
}

\email{\{pang, kawahara\}@sap.ist.i.kyoto-u.ac.jp, \{Gao\_Xiaoxue, Nancy\_Chen\}@a-star.edu.sg}


\usepackage{comment}

\begin{document}

\maketitle

\begin{abstract}

Speech emotion recognition (SER) systems can exhibit gender-related performance disparities, but how such bias manifests in multilingual speech LLMs across languages and modalities is unclear. We introduce a novel multilingual, multimodal benchmark built on MELD-ST, spanning English, Japanese, and German, to quantify language-specific SER performance and gender gaps. We find bias is strongly language-dependent, and multimodal fusion does not reliably improve fairness. To address these, we propose ERM-MinMaxGAP, a fairness-informed training objective, which augments empirical risk minimization (ERM) with a proposed adaptive fairness weight mechanism and a novel MinMaxGAP regularizer on the maximum male-female loss gap within each language and modality. Building upon the Qwen2-Audio backbone, our ERM-MinMaxGAP approach improves multilingual SER performance by 5.5\% and 5.0\% while reducing the overall gender bias gap by 0.1\% and 1.4\% in the unimodal and multimodal settings, respectively.

\end{abstract}

\begin{figure*}[htbp!]
    \centering
    \includegraphics[width=0.9\linewidth]{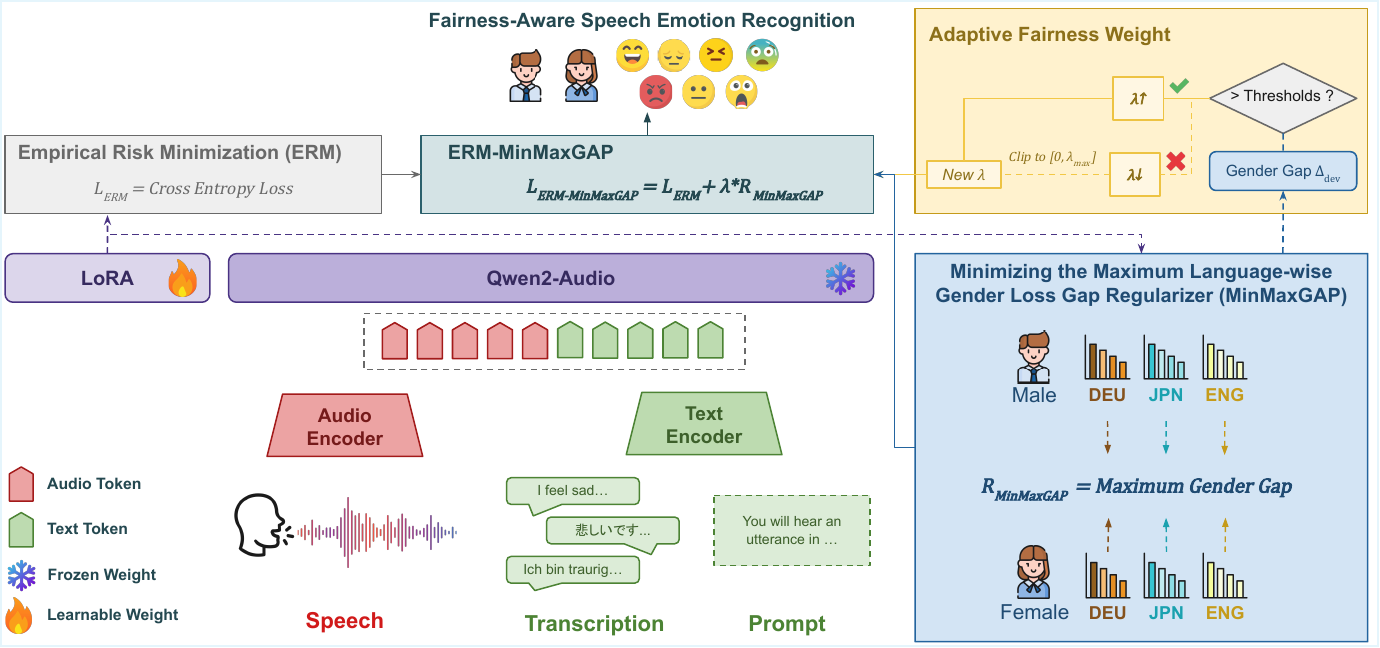}
    \caption{Architecture of the proposed method. The method consists of (1) empirical risk minimization for overall SER improvement, (2) MinMaxGAP for minimizing the language-wise gender gap, and (3) adaptive fairness-weight adjustment for fairness-aware SER.}
    \label{fig:arch}
\end{figure*}

\section{Introduction}

Speech emotion recognition (SER)~\cite{schuller2018speech} is a key capability in affective computing and spoken human--computer interaction, enabling emotion-aware conversational agents~\cite{sanjeewa2024empathic}, affective tutoring~\cite{petrovica2017emotion}, call-center analytics~\cite{martin2024speech}, and mental-health assessment~\cite{jordan2025speech,pang2026paralinguistic}. SER is commonly benchmarked on corpora such as IEMOCAP~\cite{busso2008iemocap} and MSP-IMPROV~\cite{busso2016msp}. Recent advances in self-supervised speech pretraining have substantially improved SER by transferring representations from large pretrained encoders~\cite{pepino2021wav2vec2,gao2023twostage,diatlova2024wavlm,sharma2021multilingual}. In parallel, the community is rapidly moving toward end-to-end audio--language and speech-large language models (LLMs) architectures that follow natural-language instructions and integrate audio with text, bringing SER closer to general-purpose multimodal interaction~\cite{tang2023salmonn,chu2024qwen2,rubenstein2023audiopalm,zhang2023speechgpt,bellver2024multimodal}. Despite these gains, robust deployment remains difficult due to speaker variability~\cite{sethu2013speaker,mariooryad2014compensating}, annotation subjectivity~\cite{tavernor2024annotators}, and domain mismatch across corpora and recording conditions~\cite{pastor2024domain}.

A particularly important challenge is fairness. Because emotions are expressed through acoustic--prosodic patterns that correlate with speaker traits and language, SER models may exploit demographic or linguistic shortcuts rather than emotion-relevant cues, leading to uneven performance across groups~\cite{mariooryad2014compensating,ulgen2024revealing}. Fairness issues have also been repeatedly studied in related speech technologies, including racial disparities in commercial automatic speech recognition (ASR)~\cite{koenecke2020racial}, gender and dialect bias in English ASR~\cite{tatman2017youtube,harris2024modeling}, and persistent gender performance gaps in multilingual speech recognition across many languages~\cite{attanasio2024twists,veliche2024fairspeech}. In SER specifically, prior work reported measurable gender disparities and showed that debiasing can improve consistency between female and male speakers~\cite{gorrostieta2019gender}, while more recent benchmarking indicates that mitigating disparity often involves a nontrivial fairness--accuracy trade-off, and that debiasing behavior can depend strongly on dataset composition and method choice~\cite{lin2025emo}.

However, existing SER fairness studies mainly focus on classifier-style pipelines built on fixed self-supervised learning (SSL) representations rather than modern end-to-end audio--language or speech-LLM systems~\cite{lin2025emo}. This gap matters because speech LLMs are increasingly used as general audio reasoning backbones~\cite{tang2023salmonn,chu2024qwen2,rubenstein2023audiopalm,zhang2023speechgpt}. Meanwhile, fairness analyses of speech-integrated LLMs have largely centered on semantic tasks such as speech-to-text translation, spoken coreference, sentence continuation, and spoken question answering, rather than SER~\cite{lin2024listen}. As a result, gender bias in multilingual, multimodal SER with speech LLMs remains under-benchmarked.

To fill this gap, we present a dedicated benchmark of gender bias in multilingual, multimodal speech LLMs for SER under a controlled setting that disentangles language from corpus effects. We further propose \textbf{ERM-MinMaxGAP}, a fairness-aware training objective that augments empirical risk minimization with a penalty on the maximum within-language, within-modality male--female loss gap. By explicitly penalizing the worst subgroup disparity, the objective aims to improve overall SER performance while reducing gender imbalance within each language and modality. In summary, our contributions are twofold: (1) we provide, to the best of our knowledge, the first dedicated benchmark of gender bias in multilingual multimodal speech LLMs for SER; and (2) we propose ERM-MinMaxGAP, a fairness-aware objective that minimizes overall SER loss while directly reducing the worst within-language, within-modality gender disparity. Project page: \url{https://github.com/zihaurpang/ERM-MinMaxGAP}.

\section{Methodology}

To minimize the overall gender gap in multilingual multimodal speech LLMs while preserving overall SER performance, we propose ERM-MinMaxGAP, as illustrated in Figure~\ref{fig:arch}.

\subsection{ERM-Based Supervised Fine-Tuning}

Since our main objective is to optimize SER, we start with an empirical risk minimization (ERM) procedure, i.e., minimizing the average loss over the training set. In practice, we apply standard supervised fine-tuning (SFT) with Low-Rank Adaptation (LoRA) \cite{hu2022lora} in this study. 
We use the cross-entropy loss \cite{zhang2018generalized} as our main objective function, denoted as $\mathcal{L}_{\mathrm{ERM}}(\theta) = Cross Entropy Loss$, where $\theta$ represents the model parameters, as shown in the top-left part of Figure~\ref{fig:arch}.


\subsection{MinMaxGAP Regularization}

To reduce gender disparity in multilingual SER, we introduce \textbf{Minimizing the Maximum Language-wise Gender Loss Gap Regularizer (MinMaxGAP)}, a fairness regularizer that explicitly measures the male--female loss gap \emph{within each language}, as illustrated in the bottom-right part of Figure~\ref{fig:arch}.

To measure the maximum gender gap within each language, we first define the conditional mean loss

\[
\mathcal{L}_{\ell,g}(\theta)
=
\mathbb{E}\negthinspace\left[\mathrm{CE}_i(\theta)\mid \ell_i=\ell,\; g_i=g\right],
\]


where $\ell \in \mathcal{L}$ denotes the language, $g \in \{\mathrm{F}, \mathrm{M}\}$ denotes the gender group, and \(\mathrm{CE}_i(\theta)\) denotes the per-sample cross-entropy loss. The within-language gender gap for language $\ell$ is then defined as
\[
\Delta_{\ell}(\theta)
=
\left|
\mathcal{L}_{\ell,\mathrm{F}}(\theta)
-
\mathcal{L}_{\ell,\mathrm{M}}(\theta)
\right|.
\]

To prevent a large disparity in one language from being masked by smaller disparities in others, instead of averaging the gaps across languages, we focus on the worst-case language:
\[
\Delta_{\max}(\theta)
=
\max_{\ell \in \mathcal{L}} \Delta_{\ell}(\theta).
\]

We further include a fairness regularizer that penalizes the training objective, defined as
\[
\mathcal{R}_{\mathrm{MinMaxGAP}}(\theta)
=
\left(\Delta_{\max}(\theta)\right)^p,
\]
where $p \in \{1,2\}$ controls the penalty shape. In our main setting, we use $p=2$, which places a stronger penalty on large disparities.

In practice, this regularizer is computed using mini-batch estimates of the group-wise losses. For each batch, we compute the female and male losses within each language, form the corresponding language-wise gaps, and use the maximum valid gap as the batch-level fairness signal.

\subsection{Adaptive Fairness Weight}

A fixed fairness weight is often suboptimal across training stages, especially early in training: the model should prioritize learning the fundamental SER task, whereas stronger fairness pressure may become more beneficial once the model reaches a reasonable level of performance.

To address this, we propose a dynamic fairness adjustment mechanism that adaptively updates the fairness weight based on the development-set gap.
Drawing inspiration from constrained optimization~\cite{bertsekas2014constrained}, we employ a Lagrange multiplier method to refine the weight adaptively during training (Figure~\ref{fig:arch} top right).
We update the fairness weight as
\[
\lambda^{(k+1)}
=
\Pi_{[0,\lambda_{\max}]}
\left(
\lambda^{(k)}
+
\eta\left(\Delta_{\mathrm{dev}}^{(k)} - \epsilon\right)
\right),
\]
where $\Delta_{\mathrm{dev}}^{(k)}$ denotes the fairness gap measured on the development set at evaluation step $k$, $\epsilon$ is the target tolerance, $\eta$ is the update rate, $\lambda_{\max}$ is the maximum regularization strength, and $\Pi_{[0,\lambda_{\max}]}$ denotes clipping onto the interval $[0,\lambda_{\max}]$.

Intuitively, when the observed development-set gap exceeds the target threshold $\epsilon$, the regularization weight increases, placing more emphasis on reducing disparity. When the gap is already below the threshold, the update becomes small or stops, allowing the optimization to focus more on task performance.

\subsection{Final Training Objective}

As illustrated in the top-middle part of Figure~\ref{fig:arch}, we combine the ERM term and the fairness regularizer to form the final training objective. At step $k$, the objective is defined as
\[
\mathcal{L}_{\mathrm{ERM\text{-}MinMaxGAP}}^{(k)}(\theta)
=
\mathcal{L}_{\mathrm{ERM}}(\theta)
+
\lambda^{(k)}\,\mathcal{R}_{\mathrm{MinMaxGAP}}(\theta).
\]

This objective jointly optimizes overall SER performance and worst-language gender fairness. The ERM term preserves general recognition ability, while the MinMaxGAP term directly suppresses the largest male--female loss disparity across languages.

\section{Experimental Setup}

\begin{table}[t]
\centering
\caption{Statistics of the MELD-ST dataset.}
\label{tab:meldst_stats}
\scriptsize
\begin{tabular}{llcccc}
\toprule
\textbf{Language} & \textbf{Gender} & \textbf{Train} & \textbf{Valid} & \textbf{Test} & \textbf{Total} \\
\midrule
\multirow{3}{*}{English}
 & \textit{Female} & 3546 & 464 & 501 & 4511 \\
 & \textit{Male}   & 3850 & 448 & 459 & 4757 \\
 & Total  & 7396 & 912 & 960 & 9268 \\
\midrule
\multirow{3}{*}{Japanese}
 & \textit{Female} & 3546 & 464 & 501 & 4511 \\
 & \textit{Male}   & 3850 & 448 & 459 & 4757 \\
 & Total  & 7396 & 912 & 960 & 9268 \\
\midrule
\multirow{3}{*}{German}
 & \textit{Female} & 4151 & 546 & 525 & 5222 \\
 & \textit{Male}   & 4425 & 514 & 555 & 5494 \\
 & Total  & 8576 & 1060 & 1080 & 10716 \\
\midrule
\multirow{3}{*}{Overall}
 & \textit{Female} & 11243 & 1474 & 1527 & 14244 \\
 & \textit{Male}   & 12125 & 1410 & 1473 & 15008 \\
 & Total  & 23368 & 2884 & 3000 & 29252 \\
\bottomrule
\end{tabular}
\end{table}

\subsection{Database}
We conduct experiments on \textbf{MELD-ST}~\cite{chen2024meld}, a multilingual emotion-aware speech dataset derived from MELD~\cite{poria2019meld}. It extends the original English dialogue data with aligned English--Japanese and English--German speech pairs while preserving 7-class emotion labels. To support gender-fairness analysis, we additionally annotate speaker gender. Because the released data identify speakers but do not provide explicit gender labels, we manually assign one label per speaker by cross-checking video clips, audio, and cast information. We exclude utterances with multiple speakers or unreliable gender attribution. Each remaining utterance is associated with a single speaker and a unique gender label. Final dataset statistics are shown in Table~\ref{tab:meldst_stats}.

\subsection{Training Hyperparameters}
We use \textbf{Qwen2-Audio-7B-Instruct}~\cite{chu2024qwen2} as the backbone to fine-tune on the MELD-ST. We train for up to 20 epochs with an effective batch size of 64, learning rate \(5\times10^{-5}\), weight decay 0.01, early stopping patience of 5, and random seed 42. For LoRA-based fine-tuning~\cite{hu2022lora}, we set \(r=16\), \(\alpha=32\), and dropout to 0.05. For the fairness objective, we use penalty power \(p=2\), initialize \(\lambda=0\), and enable adaptive updating with \(\epsilon=0.02\), update rate 0.5, and maximum \(\lambda=10.0\).

\subsection{Comparison Models}
We evaluate both \textbf{unimodal}, where the model receives speech and the task instruction only, and \textbf{multimodal}, where it additionally receives the ground-truth transcription. We compare our method with recent speech LLMs in the zero-shot setting: Qwen2-Audio-7B-Instruct~\cite{chu2024qwen2}, Voxtral-Mini-3B~\cite{liu2025voxtral}, gpt4o-mini-audio\footnote{\url{https://developers.openai.com/api/docs/models/gpt-4o-mini-audio-preview}}, Kimi-Audio-7B-Instruct~\cite{kimi_audio_2024}, and Ultravox-0.4\footnote{\url{https://github.com/fixie-ai/ultravox}}.

\subsection{Evaluation Metrics}
We evaluate both \textbf{SER performance} and \textbf{gender fairness}~\cite{lin2024emo, lin2025emo}. For SER, we report \textbf{weighted F1 (W-F1)} and \textbf{accuracy (ACC)}, where higher is better. For fairness, we report gender gaps in \textbf{True Positive Rate (TPR)}, \textbf{False Positive Rate (FPR)}, \textbf{W-F1}, and \textbf{ACC}. TPR/FPR gaps are computed in a one-vs-rest manner and averaged over emotion classes, while W-F1/ACC gaps measure absolute performance differences between female and male speakers. Lower values indicate smaller disparity. We further report \textbf{AVG}, the mean of the four gap metrics, as a compact summary of overall gender bias.


%
\begin{table*}[t]
\centering
\scriptsize
\caption{Comparison of unimodal and multimodal performance, with SER results and gender bias gap metrics [\%]. Relative improvements and degradations of the multimodal input over the unimodal input are shown in \textcolor{ForestGreen}{green}, and \textcolor{Maroon}{red}, respectively. Top two results are highlighted in \textbf{bold} and \underline{underline}, respectively.}
\setlength{\tabcolsep}{4pt}
\begin{tabular}{l cc ccccc @{\hspace{0.8em}}c@{\hspace{0.8em}} cc ccccc}
\toprule
\multirow{3}{*}{Model}
& \multicolumn{7}{c}{Unimodal (Speech only)}
& 
& \multicolumn{7}{c}{Multimodal (Speech + Transcription)} \\
\cmidrule(lr){2-8} \cmidrule(lr){10-16}
& \multicolumn{2}{c}{SER Result}
& \multicolumn{5}{c}{Gender Bias Gap}
&
& \multicolumn{2}{c}{SER Result}
& \multicolumn{5}{c}{Gender Bias Gap} \\
\cmidrule(lr){2-3} \cmidrule(lr){4-8} \cmidrule(lr){10-11} \cmidrule(lr){12-16}
& W-F1$\uparrow$ & ACC$\uparrow$ & TPR$\downarrow$ & FPR$\downarrow$ & W-F1$\downarrow$ & ACC$\downarrow$ & AVG$\downarrow$
&
& W-F1$\uparrow$ & ACC$\uparrow$ & TPR$\downarrow$ & FPR$\downarrow$ & W-F1$\downarrow$ & ACC$\downarrow$ & AVG$\downarrow$ \\
\midrule
\multicolumn{16}{c}{\textit{Multilingual}}
\\
\midrule
Qwen2-Audio
& 34.89 & 33.31 & 9.78 & 4.26 & 3.36 & 4.67 & 5.51
&
& 34.62 & 30.79 & 8.66 & 2.08 & 3.67 & 3.35 & 4.44 \, \textcolor{ForestGreen}{$\downarrow\negthinspace1.07$} \\

Voxtral-Mini-3B
& 44.78 & 44.52 & 9.28 & 2.74 & 4.97 & 5.22 & 5.55
&
& 50.04 & \underline{55.03} & 7.27 & \underline{1.90} & \underline{2.18} & \textbf{1.84} & \textbf{3.30} \, \textcolor{ForestGreen}{$\downarrow\negthinspace2.25$}  \\


gpt4o-mini-audio
& \underline{45.89} & \underline{44.57} & 9.61 & \textbf{1.48} & \underline{2.61} & 4.03 & 4.43
&
& \underline{52.65} & 51.76 & 8.09 & 2.08 & 4.38 & 5.25 & 4.95 \, \textcolor{Maroon}{$\uparrow\negthinspace0.52$}   \\

kimi-audio-7b
& 40.27 & 39.96 & 9.18 & 3.12 & 5.21 & \underline{3.58} & 5.27
&
& 42.34 & 42.56 & \underline{7.15} & 2.36 & 3.21 & 3.26 & 4.00 \, \textcolor{ForestGreen}{$\downarrow\negthinspace1.27$}   \\

Ultravox-0.4
& 27.43 & 25.29 & \textbf{2.76} & \underline{1.49} & \textbf{2.41} & \textbf{1.09} & \textbf{1.94}
&
& 32.45 & 30.78 & 7.79 & \textbf{1.86} & 5.65 & 4.42 & 4.93 \, \textcolor{Maroon}{$\uparrow\negthinspace2.99$}    \\

ERM-MinMaxGAP (Ours)
& \textbf{51.38} & \textbf{54.32} & \underline{6.38} & 3.01 & 3.52 & 4.44 & \underline{4.34}
&
& \textbf{57.68} & \textbf{58.65} & \textbf{7.08} & 2.69 & \textbf{1.84} & \underline{2.53} & \underline{3.53} \, \textcolor{ForestGreen}{$\downarrow\negthinspace0.80$}    \\

\midrule
\multicolumn{16}{c}{\textit{Monolingual}} \\
\midrule
\rowcolor[gray]{0.95}\multicolumn{16}{l}{\textbf{DEU}} \\

Qwen2-Audio
& 31.02 & 28.15 & 12.05 & 5.72 & 4.50 & 3.25 & 6.38
&
& 32.18 & 26.85 & 10.25 & 1.96 & 3.22 & \underline{1.49} & 4.23 \, \textcolor{ForestGreen}{$\downarrow\negthinspace2.15$} \\

Voxtral-Mini-3B
& 45.71 & \underline{45.65} & 9.60 & 2.50 & 4.09 & 4.21 & 5.10
&
& 48.46 & \underline{53.33} & 7.14 & \underline{1.47} & \underline{0.95} & \textbf{0.00} & \underline{2.39} \, \textcolor{ForestGreen}{$\downarrow\negthinspace2.71$} \\


gpt4o-mini-audio
& \underline{46.55} & 44.54 & 8.59 & \underline{1.66} & \textbf{1.60} & 3.03 & \underline{3.72}
&
& \underline{52.40} & 51.20 & 10.13 & 1.98 & 5.39 & 6.37 & 5.97 \, \textcolor{Maroon}{$\uparrow\negthinspace2.25$} \\

kimi-audio-7b
& 39.03 & 38.43 & 10.75 & 3.04 & 4.27 & \underline{2.13} & 5.05
&
& 44.16 & 42.69 & \textbf{4.14} & 1.67 & 4.53 & 5.15 & 3.87 \, \textcolor{ForestGreen}{$\downarrow\negthinspace1.17$} \\

Ultravox-0.4
& 27.97 & 25.56 & \textbf{1.68} & \textbf{0.50} & \underline{1.92} & \textbf{0.43} & \textbf{1.13}
&
& 31.71 & 28.89 & \underline{4.97} & \textbf{0.67} & \textbf{0.46} & 1.61 & \textbf{1.93} \, \textcolor{Maroon}{$\uparrow\negthinspace0.79$} \\

ERM-MinMaxGAP (Ours)
& \textbf{47.84} & \textbf{51.39} & \underline{3.06} & 3.71 & 7.47 & 9.19 & 5.86
&
& \textbf{53.32} & \textbf{54.91} & 8.01 & 3.36 & 3.92 & 4.92 & 5.05 \, \textcolor{ForestGreen}{$\downarrow\negthinspace0.81$} \\

\midrule
\rowcolor[gray]{0.95}\multicolumn{16}{l}{\textbf{ENG}} \\

Qwen2-Audio
& 46.02 & 44.27 & 7.69 & 3.65 & \underline{0.92} & 1.75 & 3.50
&
& 46.13 & 41.67 & 10.34 & 2.31 & 4.57 & 5.11 & 5.58 \, \textcolor{Maroon}{$\uparrow\negthinspace2.08$} \\

Voxtral-Mini-3B
& 41.98 & 38.75 & 9.88 & 2.75 & 7.57 & 7.87 & 7.02
&
& 53.74 & \underline{57.71} & 7.64 & 2.04 & 3.57 & 2.46 & \underline{3.93} \, \textcolor{ForestGreen}{$\downarrow\negthinspace3.09$} \\


gpt4o-mini-audio
& 49.67 & 49.06 & 12.47 & \textbf{1.52} & 4.22 & 5.93 & 6.03
&
& \underline{56.99} & 56.35 & \underline{5.77} & \textbf{1.36} & 4.42 & 5.29 & 4.21 \, \textcolor{ForestGreen}{$\downarrow\negthinspace1.82$} \\

kimi-audio-7b
& \underline{57.86} & \underline{54.79} & 10.33 & \underline{1.54} & 5.52 & 5.63 & 5.80
&
& 56.11 & 53.85 & 10.58 & \underline{1.71} & \underline{2.02} & \underline{2.17} & 4.77 \, \textcolor{ForestGreen}{$\downarrow\negthinspace0.80$} \\

Ultravox-0.4
& 35.80 & 32.50 & \textbf{4.01} & 2.43 & 1.76 & \textbf{0.49} & \textbf{2.17}
&
& 39.98 & 38.44 & 9.56 & 1.88 & 8.93 & 7.28 & 6.91 \, \textcolor{Maroon}{$\uparrow\negthinspace4.74$} \\

ERM-MinMaxGAP (Ours)
& \textbf{58.09} & \textbf{60.52} & \underline{8.84} & 2.10 & \textbf{0.41} & \underline{0.92} & \underline{3.07}
&
& \textbf{68.13} & \textbf{68.85} & \textbf{5.08} & 1.74 & \textbf{0.38} & \textbf{1.24} & \textbf{2.11} \, \textcolor{ForestGreen}{$\downarrow\negthinspace0.96$} \\

\midrule
\rowcolor[gray]{0.95}\multicolumn{16}{l}{\textbf{JPN}} \\

Qwen2-Audio
& 27.64 & 27.50 & 9.07 & 2.88 & 3.56 & 7.19 & 5.68
&
& 25.56 & 23.85 & \textbf{3.61} & \textbf{1.95} & 3.03 & 2.29 & \textbf{2.72} \, \textcolor{ForestGreen}{$\downarrow\negthinspace2.96$} \\

Voxtral-Mini-3B
& \underline{46.63} & \underline{49.17} & 8.27 & 2.96 & \underline{0.20} & \textbf{1.39} & 3.21
&
& 47.91 & \textbf{54.06} & 7.01 & \underline{2.12} & \textbf{0.81} & 2.03 & 2.99 \, \textcolor{ForestGreen}{$\downarrow\negthinspace0.22$} \\


gpt4o-mini-audio
& 41.45 & 40.10 & 6.89 & \underline{1.24} & \textbf{0.11} & 2.12 & \underline{2.59}
&
& \underline{48.56} & 47.71 & 7.77 & 2.69 & 2.99 & 3.75 & 4.30 \, \textcolor{Maroon}{$\uparrow\negthinspace1.71$} \\

kimi-audio-7b
& 23.91 & 26.67 & \underline{5.50} & 4.18 & 5.72 & \underline{1.50} & 4.23
&
& 26.76 & 31.15 & \underline{4.95} & 3.32 & 2.52 & \textbf{0.82} & \underline{2.90} \, \textcolor{ForestGreen}{$\downarrow\negthinspace1.32$} \\

Ultravox-0.4
& 18.52 & 17.81 & \textbf{2.01} & \textbf{0.74} & 3.25 & 1.77 & \textbf{1.94}
&
& 25.66 & 25.00 & 8.12 & 2.52 & 3.99 & 1.77 & 4.10 \, \textcolor{Maroon}{$\uparrow\negthinspace2.16$} \\

ERM-MinMaxGAP (Ours)
& \textbf{48.19} & \textbf{51.04} & 7.23 & 3.21 & 2.68 & 3.22 & 4.09
&
& \textbf{51.58} & \underline{52.19} & 8.13 & 2.97 & \underline{1.22} & \underline{1.44} & 3.44 \, \textcolor{ForestGreen}{$\downarrow\negthinspace0.64$} \\

\bottomrule
\end{tabular}

\label{tab:ser_gender_bias_comparison}
\end{table*}
\begin{table*}[htbp!]
\centering
\scriptsize
\caption{Ablation study [\%]. Relative improvements and degradations of the multimodal input over the unimodal input are shown in \textcolor{ForestGreen}{green}, and \textcolor{Maroon}{red}, respectively.}
\setlength{\tabcolsep}{4pt}
\begin{tabular}{l cc ccccc @{\hspace{0.8em}}c@{\hspace{0.8em}} cc ccccc}
\toprule
\multirow{3}{*}{Model}
& \multicolumn{7}{c}{Unimodal (Speech only)}
&
& \multicolumn{7}{c}{Multimodal (Speech + Transcription)} \\
\cmidrule(lr){2-8} \cmidrule(lr){10-16}
& \multicolumn{2}{c}{SER Result}
& \multicolumn{5}{c}{Gender Bias Gap}
&
& \multicolumn{2}{c}{SER Result}
& \multicolumn{5}{c}{Gender Bias Gap} \\
\cmidrule(lr){2-3} \cmidrule(lr){4-8} \cmidrule(lr){10-11} \cmidrule(lr){12-16}
& W-F1$\uparrow$ & ACC$\uparrow$ & TPR$\downarrow$ & FPR$\downarrow$ & W-F1$\downarrow$ & ACC$\downarrow$ & AVG$\downarrow$
&
& W-F1$\uparrow$ & ACC$\uparrow$ & TPR$\downarrow$ & FPR$\downarrow$ & W-F1$\downarrow$ & ACC$\downarrow$ & AVG$\downarrow$ \\
\midrule
\rowcolor[gray]{0.95}\multicolumn{16}{l}{\textbf{Main Components}} \\

Qwen2-Audio
& 34.89 & 33.31 & 9.78 & 4.26 & 3.36 & 4.67 & 5.51
&
& 34.62 & 30.79 & 8.66 & 2.08 & 3.67 & 3.35 & 4.44 \, \textcolor{ForestGreen}{$\downarrow\negthinspace1.07$} \\

+ ERM (SFT)
& 47.50 & 46.07 & 11.28 & 2.64 & 2.87 & 3.08 & 4.97
&
& 56.13 & 54.77 & 9.50 & 2.03 & 0.93 & 1.22 & 3.42 \, \textcolor{ForestGreen}{$\downarrow\negthinspace1.55$} \\

ERM-MinMaxGAP (Ours)
& 51.38 & 54.32 & 6.38 & 3.01 & 3.52 & 4.44 & 4.34
&
& 57.68 & 58.65 & 7.08 & 2.69 & 1.84 & 2.53 & 3.53 \, \textcolor{ForestGreen}{$\downarrow\negthinspace0.80$} \\

\midrule
\rowcolor[gray]{0.95}\multicolumn{16}{l}{\textbf{Lambda Effect}} \\

$\lambda = 0$ (SFT)
& 47.50 & 46.07 & 11.28 & 2.64 & 2.87 & 3.08 & 4.97
&
& 56.13 & 54.77 & 9.50 & 2.03 & 0.93 & 1.22 & 3.42 \, \textcolor{ForestGreen}{$\downarrow\negthinspace1.55$} \\

$\lambda = 1$
& 32.71 & 37.86 & 3.31 & 1.59 & 5.73 & 4.33 & 3.74
&
& 47.61 & 45.81 & 7.62 & 1.38 & 1.17 & 1.59 & 2.94 \, \textcolor{ForestGreen}{$\downarrow\negthinspace0.80$} \\

$\lambda = 5$
& 30.97 & 30.17 & 3.83 & 0.72 & 3.33 & 3.15 & 2.76
&
& 35.50 & 33.31 & 6.04 & 1.31 & 2.03 & 2.12 & 2.87 \, \textcolor{Maroon}{$\uparrow\negthinspace0.12$}  \\

$\lambda = 10$
& 29.67 & 28.77 & 3.87 & 1.23 & 2.23 & 2.55 & 2.47
&
& 30.87 & 31.04 & 3.12 & 0.68 & 2.92 & 3.35 & 2.52 \, \textcolor{Maroon}{$\uparrow\negthinspace0.05$} \\

$\lambda = \mathrm{adaptive}$ (Ours)
& 51.38 & 54.32 & 6.38 & 3.01 & 3.52 & 4.44 & 4.34
&
& 57.68 & 58.65 & 7.08 & 2.69 & 1.84 & 2.53 & 3.53 \, \textcolor{ForestGreen}{$\downarrow\negthinspace0.80$} \\

\midrule
\rowcolor[gray]{0.95}\multicolumn{16}{l}{\textbf{Penalty Power Effect}} \\

$p = 1$
& 51.38 & 54.32 & 6.38 & 3.01 & 3.52 & 4.44 & 4.34
&
& 58.76 & 59.77 & 7.91 & 2.81 & 1.91 & 2.30 & 3.73 \, \textcolor{ForestGreen}{$\downarrow\negthinspace0.61$} \\

$p = 2$ (Ours)
& 51.38 & 54.32 & 6.38 & 3.01 & 3.52 & 4.44 & 4.34
&
& 57.68 & 58.65 & 7.08 & 2.69 & 1.84 & 2.53 & 3.53 \, \textcolor{ForestGreen}{$\downarrow\negthinspace0.80$} \\
\bottomrule
\end{tabular}
\label{tab:ablation_study}
\end{table*}

\section{Results and Analysis}

\subsection{Benchmarking Gender Bias in Multilingual Multimodal Speech LLMs}

Table~\ref{tab:ser_gender_bias_comparison} benchmarks gender bias together with overall SER performance for recent speech LLMs across languages and modalities. We find that multimodal input often improves SER, but does not consistently improve fairness. Some models (e.g., Voxtral-Mini-3B and kimi-audio-7b) show gains in both SER and reduced gender gap, whereas others improve overall SER while exhibiting a larger gender gap, indicating that multimodal fusion can improve recognition accuracy without reliably reducing gender disparity.

The benchmark also shows strong language dependence. English is generally the easiest setting and Japanese the most difficult, with stronger models achieving higher W-F1/ACC in ENG than in JPN under both input conditions. Fairness trends are likewise inconsistent across languages: a model may reduce the gender gap in one language but enlarge it in another. Overall, gender bias in multilingual multimodal SER is highly model- and language-dependent, motivating a fairness-aware objective rather than relying on multimodal fusion alone.
\subsection{Effectiveness of ERM-MinMaxGAP}

As shown in Table~\ref{tab:ser_gender_bias_comparison}, \textbf{ERM-MinMaxGAP} achieves the best overall performance while consistently reducing gender disparity. In the multilingual setting, it obtains the best SER in both conditions, with gains of +5.49 W-F1 and +9.75 ACC in the unimodal setting, and +5.03 W-F1 and +3.62 ACC in the multimodal setting, relative to the best baseline, while maintaining the second-smallest gender gap. Importantly, its AVG gender gap is also reduced by 0.8 when moving to multimodal input. 

In the monolingual settings, ERM-MinMaxGAP remains the strongest overall performer in all languages while keeping gender disparity as small as possible. It is also observed that it provides consistent mitigation of the AVG gender gap in each language. These results show that ERM-MinMaxGAP improves SER while delivering a stronger overall performance--fairness trade-off across languages and modalities. Although it does not always yield the minimum post-hoc gap in every setting, this is consistent with its goal of penalizing the worst within-language group disparity rather than directly minimizing an average fairness metric.

\subsection{Ablation Study}

Table~\ref{tab:ablation_study} confirms the contribution of each component to \textbf{ERM-MinMaxGAP}. Relative to zero-shot Qwen2-Audio, ERM-based supervised fine-tuning (SFT) already provides a large gain while reducing the AVG gender gap in both the unimodal and multimodal settings. Moreover, our proposed MinMaxGAP further improves overall SER performance while reducing gender disparity in both settings. This suggests that MinMaxGAP improves the overall performance--fairness trade-off rather than uniformly minimizing every post-hoc gap metric. We also evaluated a MinMaxGAP-only ablation, but without the main task objective (ERM), the model could not produce a valid single prediction; therefore, this setting was excluded from the study.

We also evaluated the effect of the fairness weight $\lambda$. Compared with our proposed adaptive weight update method, a fixed $\lambda$ reduces the gender gap as it increases, but also degrades SER, revealing a clear fairness--utility trade-off. We also examined whether the penalty power affects overall performance. We find that although $p=1$ yields slightly higher multimodal SER, it also produces a larger gender gap, whereas our proposed $p=2$ provides better fairness.

\section{Conclusion}

This paper presented, to the best of our knowledge, the first dedicated benchmark of gender bias in multilingual multimodal speech LLMs for SER. Through evaluation on MELD-ST across English, Japanese, and German, we showed that gender disparity is highly dependent on both language and input modality, and that multimodal input does not reliably reduce bias even when it improves SER performance. To address this issue, we proposed \textbf{ERM-MinMaxGAP}, a fairness-aware training objective that combines empirical risk minimization with worst-language gender-gap regularization. Experimental results showed that ERM-MinMaxGAP achieves the strongest overall performance while providing more consistent gender-gap mitigation across languages and modalities, yielding a better performance--fairness trade-off than the compared baselines. 


\section{Generative AI Use Disclosure}

Generative AI tools were used only for language editing and polishing to improve the clarity and readability of this manuscript. They were not used to generate scientific content, research ideas, experimental designs, results, or conclusions. All content was reviewed by the authors, who take full responsibility for the manuscript.

\section{Acknowledgment}

This research was supported by Japan-Singapore Joint Call: Japan Science and Technology Agency (JST) (No. 251043539) and A*STAR 2024 (R24I6IR136), the National Research Foundation,
Singapore under its National Large Language Models Funding Initiative (AISG Award No: AISG-NMLP-2024-004), and the National Research Foundation, Singapore under its National Large Language Models Funding Initiative (AISG Award No: AISG-NMLP-2024-003). Any opinions, findings and conclusions or recommendations expressed in this material are those of the author(s) and do not reflect the views of National Research Foundation, Singapore.

\bibliographystyle{IEEEtran}
\bibliography{mybib}

@inproceedings{gorrostieta2019gender,
  title={Gender De-Biasing in Speech Emotion Recognition.},
  author={Gorrostieta, Cristina and Lotfian, Reza and Taylor, Kye and Brutti, Richard and Kane, John},
  booktitle={Interspeech},
  pages={2823--2827},
  year={2019}
}

@inproceedings{chen2024meld,
  title={MELD-ST: An emotion-aware speech translation dataset},
  author={Chen, Sirou and Yahata, Sakiko and Shimizu, Shuichiro and Yang, Zhengdong and Li, Yihang and Chu, Chenhui and Kurohashi, Sadao},
  booktitle={Findings of the Association for Computational Linguistics: ACL 2024},
  pages={10118--10126},
  year={2024}
}

@inproceedings{poria2019meld,
  title={Meld: A multimodal multi-party dataset for emotion recognition in conversations},
  author={Poria, Soujanya and Hazarika, Devamanyu and Majumder, Navonil and Naik, Gautam and Cambria, Erik and Mihalcea, Rada},
  booktitle={Proceedings of the 57th annual meeting of the association for computational linguistics},
  pages={527--536},
  year={2019}
}

@article{sanjeewa2024empathic,
  title={Empathic conversational agent platform designs and their evaluation in the context of mental health: systematic review},
  author={Sanjeewa, Ruvini and Iyer, Ravi and Apputhurai, Pragalathan and Wickramasinghe, Nilmini and Meyer, Denny},
  journal={JMIR Mental Health},
  volume={11},
  pages={e58974},
  year={2024},
  publisher={JMIR Publications Toronto, Canada}
}

@article{petrovica2017emotion,
  title={Emotion recognition in affective tutoring systems: Collection of ground-truth data},
  author={Petrovica, Sintija and Anohina-Naumeca, Alla and Ekenel, Haz{\i}m Kemal},
  journal={Procedia Computer Science},
  volume={104},
  pages={437--444},
  year={2017},
  publisher={Elsevier}
}

@inproceedings{martin2024speech,
  title={Speech emotion recognition for call centers using self-supervised models: A complete pipeline for industrial applications},
  author={Mart{\'\i}n-Do{\~n}as, Juan M and Zorrilla, Asier L{\'o}pez and deVelasco, Mikel and V{\'a}squez-Correa, Juan Camilo and {\'A}lvarez, Aitor and Torres, Maria In{\'e}s and Delgado, Paz and Lazpiur, Ane and Romero, Blanca and Alkorta, Irati},
  booktitle={Proceedings of the 7th International Conference on Natural Language and Speech Processing (ICNLSP 2024)},
  pages={119--128},
  year={2024}
}

@article{jordan2025speech,
  title={Speech emotion recognition in mental health: Systematic review of voice-based applications},
  author={Jordan, Eric and Terrisse, Rapha{\"e}l and Lucarini, Valeria and Alrahabi, Motasem and Krebs, Marie-Odile and Descl{\'e}s, Julien and Lemey, Christophe},
  journal={JMIR mental health},
  volume={12},
  number={1},
  pages={e74260},
  year={2025},
  publisher={JMIR Publications Inc., Toronto, Canada}
}

@article{bellver2024multimodal,
  title={Multimodal audio-language model for speech emotion recognition},
  author={Bellver Soler, Jaime and Mart{\'\i}n Fern{\'a}ndez, Iv{\'a}n and Bravo Pacheco, Jose Manuel and Esteban Romero, Sergio and Fern{\'a}ndez Mart{\'\i}nez, Fernando and D'Haro Enr{\'\i}quez, Luis Fernando},
  journal={The Speaker and Language Recognition Workshop (Odyssey 2024)},
  year={2024},
}

@article{lin2024emo,
  title={Emo-bias: A large scale evaluation of social bias on speech emotion recognition},
  author={Lin, Yi-Cheng and Wu, Haibin and Chou, Huang-Cheng and Lee, Chi-Chun and Lee, Hung-yi},
  journal={INTERSPEECH 2024},
  year={2024}
}

@article{lin2025emo,
  title={Emo-debias: Benchmarking gender debiasing techniques in multi-label speech emotion recognition},
  author={Lin, Yi-Cheng and Chou, Huang-Cheng and Liang, Yu-Hsuan Li and Lee, Hung-yi},
  journal={IEEE Automatic Speech Recognition and Understanding Workshop (ASRU 2025)},
  year={2025}
}

@article{chu2024qwen2,
  title={Qwen2-audio technical report},
  author={Chu, Yunfei and Xu, Jin and Yang, Qian and Wei, Haojie and Wei, Xipin and Guo, Zhifang and Leng, Yichong and Lv, Yuanjun and He, Jinzheng and Lin, Junyang and others},
  journal={arXiv preprint arXiv:2407.10759},
  year={2024}
}

@article{tang2023salmonn,
  title={Salmonn: Towards generic hearing abilities for large language models},
  author={Tang, Changli and Yu, Wenyi and Sun, Guangzhi and Chen, Xianzhao and Tan, Tian and Li, Wei and Lu, Lu and Ma, Zejun and Zhang, Chao},
  journal={The Twelfth International Conference on Learning Representations},
  year={2024}
}

@inproceedings{lin2024listen,
  title={Listen and speak fairly: a study on semantic gender bias in speech integrated large language models},
  author={Lin, Yi-Cheng and Lin, Tzu-Quan and Yang, Chih-Kai and Lu, Ke-Han and Chen, Wei-Chih and Kuan, Chun-Yi and Lee, Hung-yi},
  booktitle={2024 IEEE Spoken Language Technology Workshop (SLT)},
  pages={439--446},
  year={2024},
  organization={IEEE}
}

@inproceedings{attanasio2024twists,
  title={Twists, humps, and pebbles: Multilingual speech recognition models exhibit gender performance gaps},
  author={Attanasio, Giuseppe and Savoldi, Beatrice and Fucci, Dennis and Hovy, Dirk},
  booktitle={Proceedings of the 2024 Conference on Empirical Methods in Natural Language Processing},
  pages={21318--21340},
  year={2024}
}

@article{koenecke2020racial,
  title={Racial disparities in automated speech recognition},
  author={Koenecke, Allison and Nam, Andrew and Lake, Emily and Nudell, Joe and Quartey, Minnie and Mengesha, Zion and Toups, Connor and Rickford, John R and Jurafsky, Dan and Goel, Sharad},
  journal={Proceedings of the national academy of sciences},
  volume={117},
  number={14},
  pages={7684--7689},
  year={2020},
  publisher={National Academy of Sciences}
}

@inproceedings{harris2024modeling,
  title={Modeling gender and dialect bias in automatic speech recognition},
  author={Harris, Camille and Mgbahurike, Chijioke and Kumar, Neha and Yang, Diyi},
  booktitle={Findings of the Association for Computational Linguistics: EMNLP 2024},
  pages={15166--15184},
  year={2024}
}

@article{hu2022lora,
  title={Lora: Low-rank adaptation of large language models.},
  author={Hu, Edward J and Shen, Yelong and Wallis, Phillip and Allen-Zhu, Zeyuan and Li, Yuanzhi and Wang, Shean and Wang, Liang and Chen, Weizhu and others},
  journal={Iclr},
  volume={1},
  number={2},
  pages={3},
  year={2022}
}

@misc{kimi_audio_2024,
      title={Kimi-Audio Technical Report},
      author={Kimi Team},
      year={2024},
      eprint={arXiv:placeholder},
      archivePrefix={arXiv},
      primaryClass={cs.CL}
}

@article{liu2025voxtral,
  title={Voxtral},
  author={Liu, Alexander H and Ehrenberg, Andy and Lo, Andy and Denoix, Cl{\'e}ment and Barreau, Corentin and Lample, Guillaume and Delignon, Jean-Malo and Chandu, Khyathi Raghavi and von Platen, Patrick and Muddireddy, Pavankumar Reddy and others},
  journal={arXiv preprint arXiv:2507.13264},
  year={2025}
}

@article{schuller2018speech,
  title={Speech emotion recognition: Two decades in a nutshell, benchmarks, and ongoing trends},
  author={Schuller, Bj{\"o}rn W},
  journal={Communications of the ACM},
  volume={61},
  number={5},
  pages={90--99},
  year={2018},
  publisher={ACM New York, NY, USA}
}

@article{mariooryad2014compensating,
  title={Compensating for speaker or lexical variabilities in speech for emotion recognition},
  author={Mariooryad, Soroosh and Busso, Carlos},
  journal={Speech Communication},
  volume={57},
  pages={1--12},
  year={2014},
  publisher={Elsevier}
}

@inproceedings{ulgen2024revealing,
  title={Revealing emotional clusters in speaker embeddings: A contrastive learning strategy for speech emotion recognition},
  author={Ulgen, Ismail Rasim and Du, Zongyang and Busso, Carlos and Sisman, Berrak},
  booktitle={ICASSP 2024-2024 IEEE International Conference on Acoustics, Speech and Signal Processing (ICASSP)},
  pages={12081--12085},
  year={2024},
  organization={IEEE}
}

@inproceedings{pang2026paralinguistic,
  title={Paralinguistic Emotion-Aware Validation Timing Detection in Japanese Empathetic Spoken Dialogue},
  author={Pang, Zi Haur and Fu, Yahui and Gao, Yuan and Kawahara, Tatsuya},
  booktitle={ICASSP 2026-2026 IEEE International Conference on Acoustics, Speech and Signal Processing (ICASSP)},
  pages={19417--19421},
  year={2026},
  organization={IEEE}
}

@article{busso2008iemocap,
  title={IEMOCAP: Interactive emotional dyadic motion capture database},
  author={Busso, Carlos and Bulut, Murtaza and Lee, Chi-Chun and Kazemzadeh, Abe and Mower, Emily and Kim, Samuel and Chang, Jeannette N and Lee, Sungbok and Narayanan, Shrikanth S},
  journal={Language resources and evaluation},
  volume={42},
  number={4},
  pages={335--359},
  year={2008},
  publisher={Springer}
}

@article{busso2016msp,
  title={MSP-IMPROV: An acted corpus of dyadic interactions to study emotion perception},
  author={Busso, Carlos and Parthasarathy, Srinivas and Burmania, Alec and AbdelWahab, Mohammed and Sadoughi, Najmeh and Provost, Emily Mower},
  journal={IEEE transactions on affective computing},
  volume={8},
  number={1},
  pages={67--80},
  year={2016},
  publisher={IEEE}
}

@book{bertsekas2014constrained,
  title={Constrained optimization and Lagrange multiplier methods},
  author={Bertsekas, Dimitri P},
  year={2014},
  publisher={Academic press}
}

@inproceedings{pepino2021wav2vec2,
  title     = {{Emotion Recognition from Speech Using wav2vec 2.0 Embeddings}},
  author    = {Leonardo Pepino and Pablo Riera and Luciana Ferrer},
  year      = {2021},
  booktitle = {{Interspeech 2021}},
  pages     = {3400--3404},
  doi       = {10.21437/Interspeech.2021-703},
  issn      = {2958-1796},
}

@inproceedings{gao2023twostage,
  title     = {{Two-stage Finetuning of Wav2vec 2.0 for Speech Emotion Recognition with ASR and Gender Pretraining}},
  author    = {Yuan Gao and Chenhui Chu and Tatsuya Kawahara},
  year      = {2023},
  booktitle = {{Interspeech 2023}},
  pages     = {3637--3641},
  doi       = {10.21437/Interspeech.2023-756},
  issn      = {2958-1796},
}

@article{diatlova2024wavlm,
  title={Adapting wavlm for speech emotion recognition},
  author={Diatlova, Daria and Udalov, Anton and Shutov, Vitalii and Spirin, Egor},
  journal={The Speaker and Language Recognition Workshop (Odyssey 2024)},
  year={2024}
}

@inproceedings{sharma2021multilingual,
  title={Multi-lingual multi-task speech emotion recognition using wav2vec 2.0},
  author={Sharma, Mayank},
  booktitle={ICASSP 2022-2022 IEEE international conference on acoustics, speech and signal processing (ICASSP)},
  pages={6907--6911},
  year={2022},
  organization={IEEE}
}

@article{rubenstein2023audiopalm,
  title={Audiopalm: A large language model that can speak and listen},
  author={Rubenstein, Paul K and Asawaroengchai, Chulayuth and Nguyen, Duc Dung and Bapna, Ankur and Borsos, Zal{\'a}n and Quitry, F{\'e}lix de Chaumont and Chen, Peter and Badawy, Dalia El and Han, Wei and Kharitonov, Eugene and others},
  journal={arXiv preprint arXiv:2306.12925},
  year={2023}
}

@inproceedings{zhang2023speechgpt,
    title = "{S}peech{GPT}: Empowering Large Language Models with Intrinsic Cross-Modal Conversational Abilities",
    author = "Zhang, Dong  and
      Li, Shimin  and
      Zhang, Xin  and
      Zhan, Jun  and
      Wang, Pengyu  and
      Zhou, Yaqian  and
      Qiu, Xipeng",
    editor = "Bouamor, Houda  and
      Pino, Juan  and
      Bali, Kalika",
    booktitle = "Findings of the Association for Computational Linguistics: EMNLP 2023",
    month = dec,
    year = "2023",
    address = "Singapore",
    publisher = "Association for Computational Linguistics",
    pages = "15757--15773",

}

@inproceedings{tatman2017youtube,
    title = "Gender and Dialect Bias in {Y}ou{T}ube{'}s Automatic Captions",
    author = "Tatman, Rachael",
    editor = "Hovy, Dirk  and
      Spruit, Shannon  and
      Mitchell, Margaret  and
      Bender, Emily M.  and
      Strube, Michael  and
      Wallach, Hanna",
    booktitle = "Proceedings of the First {ACL} Workshop on Ethics in Natural Language Processing",
    month = apr,
    year = "2017",
    address = "Valencia, Spain",
    publisher = "Association for Computational Linguistics",
    pages = "53--59",
}

@article{veliche2024fairspeech,
  title={Towards measuring fairness in speech recognition: Fair-speech dataset},
  author={Veliche, Irina-Elena and Huang, Zhuangqun and Kochaniyan, Vineeth Ayyat and Peng, Fuchun and Kalinli, Ozlem and Seltzer, Michael L},
  journal={Interspeech 2024},
  year={2024}
}

@article{tavernor2024annotators,
  title={The whole is bigger than the sum of its parts: Modeling individual annotators to capture emotional variability},
  author={Tavernor, James and El-Tawil, Yara and Provost, Emily Mower},
  journal={Interspeech 2024},
  year={2024}
}

@inproceedings{sethu2013speaker,
  title={Speaker variability in speech based emotion models-Analysis and normalisation},
  author={Sethu, Vidhyasaharan and Epps, Julien and Ambikairajah, Eliathamby},
  booktitle={2013 IEEE International Conference on Acoustics, Speech and Signal Processing},
  pages={7522--7526},
  year={2013},
  organization={IEEE}
}

@inproceedings{pastor2024domain,
  title={Analysis of the domain mismatch problem in the Speech Emotion Recognition Task},
  author={Pastor, Miguel A and Ortega, Alfonso and Ribas, Dayana},
  booktitle={Proc. IberSPEECH 2024},
  pages={181--185},
  year={2024}
}

@article{zhang2018generalized,
  title={Generalized cross entropy loss for training deep neural networks with noisy labels},
  author={Zhang, Zhilu and Sabuncu, Mert},
  journal={Advances in neural information processing systems},
  volume={31},
  year={2018}
}

\end{document}